\documentclass[twocolumn]{IEEEtran}
\usepackage[T1]{fontenc}
\usepackage[latin9]{inputenc}
\usepackage{color}
\usepackage{array}
\usepackage{multirow}
\usepackage{graphicx}
\usepackage[unicode=true,
 bookmarks=true,bookmarksnumbered=true,bookmarksopen=true,bookmarksopenlevel=1,
 breaklinks=false,pdfborder={0 0 0},pdfborderstyle={},backref=false,colorlinks=false]
 {hyperref}
\hypersetup{pdftitle={Your Title},
 pdfauthor={Your Name},
 pdfpagelayout=OneColumn,pdfnewwindow=true, pdfstartview=XYZ, plainpages=false}
\usepackage{breakurl}

\makeatletter

\providecommand{\tabularnewline}{\\}

\usepackage[caption=false,font=footnotesize]{subfig}
\usepackage{algorithm}
\usepackage{algorithmic}

\usepackage{multirow} 
\usepackage{amsmath} 
\usepackage{xcolor}

\allowdisplaybreaks[4]

\ifCLASSOPTIONcompsoc
\usepackage[caption=false,font=normalsize,labelfont=sf,textfont=sf]{subfig}
\else
\usepackage[caption=false,font=footnotesize]{subfig}
\fi

\usepackage{cite}
\usepackage{bm}
\usepackage{algorithmic}
\usepackage{algorithm}
\usepackage{graphicx}
\IEEEoverridecommandlockouts

\usepackage{lettrine}

\makeatother

\begin{document}

\title{\textcolor{black}{The Potential of LEO Satellites in 6G Space-Air-Ground
Enabled Access Networks}}

\author{Ziye Jia, \IEEEmembership{Member,~IEEE}, Chao Dong, \IEEEmembership{Member,~IEEE},
Kun Guo, \IEEEmembership{Member,~IEEE}, and \\ Qihui Wu, \IEEEmembership{Senior Member,~IEEE}\thanks{Ziye Jia is with the College of Electronic and Information Engineering,
Nanjing University of Aeronautics and Astronautics, Nanjing 211106,
China, and also with the State Key Laboratory of ISN, Xidian University,
Xi\textquoteright an 710071, China (e-mail: jiaziye@nuaa.edu.cn).

Chao Dong and Qhui Wu are with the College of Electronic and Information
Engineering, Nanjing University of Aeronautics and Astronautics, Nanjing
211106, China (e-mail: wuqihui@nuaa.edu.cn, dch@nuaa.edu.cn).

\textit{\textcolor{black}{}}Kun Guo is with the East China Normal
University, Shanghai 200241, China (e-mail: kguo@cee.ecnu.edu.cn).}}
\maketitle
\begin{abstract}
Space-air-ground integrated networks (SAGINs) help enhance the service
performance in the sixth generation communication system. SAGIN is
basically composed of satellites, aerial vehicles, ground facilities,
as well as multiple terrestrial users. Therein, the low earth orbit
(LEO) satellites are popular in recent years due to the low cost of
development and launch, global coverage and delay-enabled services.
Moreover, LEO satellites can support various applications, e.g., direct
access, relay, caching and computation. In this work, we firstly provide
the preliminaries and framework of SAGIN, in which the characteristics
of LEO satellites, high altitude platforms, as well as unmanned aerial
vehicles are analyzed. Then, the roles and potentials of LEO satellite
in SAGIN are analyzed for access services. A couple of advanced techniques
such as multi-access edge computing (MEC) and network function virtualization
are introduced to enhance the LEO-based access service abilities as
hierarchical MEC and network slicing in SAGIN. In addition,  corresponding
use cases are provided to verify the propositions. Besides, we also
discuss the open issues and promising directions in LEO-enabled SAGIN
access services for the future research.
\end{abstract}

\begin{IEEEkeywords}
Space-air-ground integrated network (SAGIN), low earth orbit (LEO)
satellite, multi-access edge computing (MEC), network slicing.
\end{IEEEkeywords}

\section{Introduction}

\subsection{Background and Motivation}

The sixth generation communication system (6G) related researches
have gained great interests from both industries and academics, among
which the space-air-ground integrated network (SAGIN) is a promising
technique to extend the range of global services and worldwide Internet
access \cite{1263-xiaozhenyu-JSAC2022,jzy-IoT-bender,1264-xiaozhenyu-CC2022}.
Specifically, with the explosive growth of satellite businesses from
various individuals and institutions, such as satellite navigation,
emergency rescue, and deep space exploration, there exists significant
demands to enhance the performance of traditional communication networks
\cite{1268-helijun-TMC2}, and SAGIN related techniques are promising
to satisfy such requirements. Besides, in order to enable the region
coverage and access quality, with the limitation of ground infrastructures,
SAGIN enabled access and global coverage are significant.

\begin{figure*}
\centering

\includegraphics[scale=0.27]{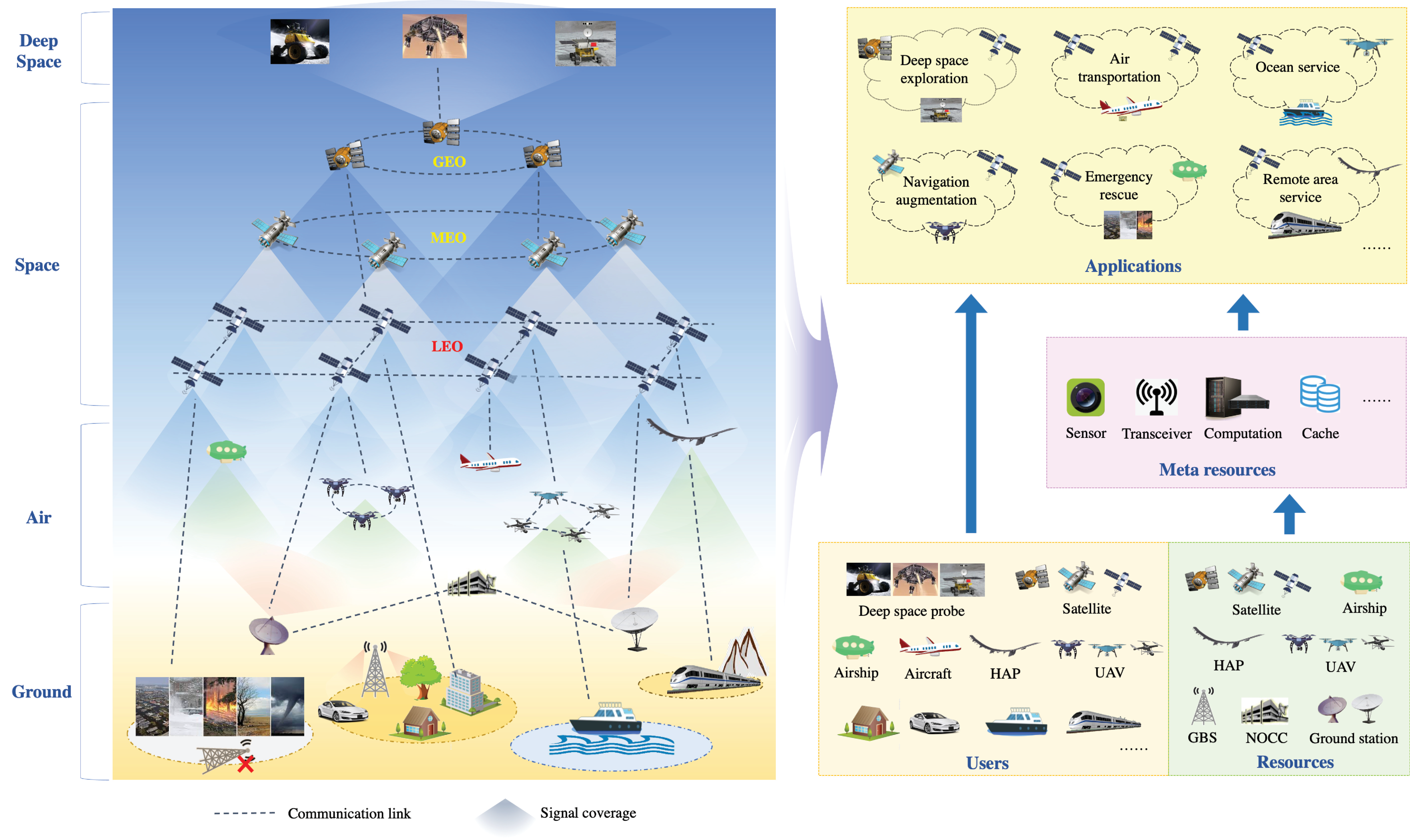}\textcolor{black}{\caption{\textcolor{black}{\label{fig:SAGIN}A view of SAGIN scenario.}}
}
\end{figure*}

In particular, SAGIN is typically composed of three layers according
to the altitude, i.e., space satellites, aerial vehicles, and terrestrial
facilities, as the illustration in Fig. \ref{fig:SAGIN}. Specifically,
in the space, satellites are basically divided into three types based
on different heights, i.e., geosynchronous earth orbit (GEO) satellites,
medium earth orbit (MEO) satellites, and low earth orbit (LEO) satellites
\cite{1260-xiaozhenyu-WC,1273-xuelin-JSAC-LEO}. The high altitude
platforms (HAPs) and unmanned air vehicles (UAVs) are popular facilities
in the air \cite{ziyejia-jsac,1020-dongchao-mag-uav}. HAPs are characterized
for the stable and large coverage, while UAVs are specialized in flexible
services \cite{990-HAP-mag2021}. Besides, there exists heterogenous
resources such as sensors, transmitters, computation modules, caches,
etc. Among which, LEO satellites play prominent roles in terms of
networking and constellations, and we mostly focus on elaborating
the roles and potentials of LEO satellites in the SAGIN access networks.

However, there are multiple heterogenous users such as remote Internet
of things (IoT), maritime voyages, and air transportation. Furthermore,
the number of such users has a sharp increment in recent years. \textcolor{black}{It
is challenging to satisfy the ever-increasing requirements.} However,
the number of platforms in SAGIN, as well as the corresponding loading
capacity such as \textcolor{black}{hardwares,} are limited\textcolor{black}{.
Besides, the resources in SAGIN are dynamic and the operational modes
of various platforms are quite diverse. For example, LEO satellites
in different orbits have different cycle periods, the movement of
UAV is flexible but unpredictable, while HAPs are quasi-static. Also,
the scarce intermittent communication resources caused by the periodic
motion of satellites aggravate the resource competition.} Consequently,
how to figure out such \textcolor{black}{challenges, i.e., heterogeneous
platforms, various demands with diverse quality of service (QoS) requirements,
and}\textcolor{blue}{{} }\textcolor{black}{limited resources, are significant
issues. }

As above, an elastic and scalable SAGIN should be well designed to
satisfy the various demands. Besides, the potential of LEO satellites
should be explored due to the increasing number of various LEO satellite
constellation, such as Starlink \cite{1276-role-LEO}.\textcolor{black}{{}
The advanced }in-network computing paradigm\textcolor{black}{s such
as multi-access edge computing (MEC), network function virtualization
(NFV), and software defined network (SDN), are introduced into the
SAGIN for efficient and resilient resource providers \cite{JZY-TWC}.
SDN can assist manage SAGIN by the pattern of separating the control
plane and data plane.} NFV helps to provide multiple services for
different users via decoupling the virtual network functions (VNFs)
from physical platforms. In addition, a serious of VNFs for one task
form a service function chain (SFC). The MEC paradigm can be utilized
to help the remote users in SAGIN for computation task offloading
\cite{350-SapceIOTml-chengnanjsac}. \textcolor{black}{In short, MEC,
NFV, and SDN are }prospective modes for flexible resource provider,
and can help deal with multiple tasks with heterogeneous platforms
and limited resources in SAGIN.

\subsection{State of the Art}

The SAGIN as well as LEO satellites, are widely studied recently.
For example, in \cite{1268-helijun-TMC2}\textcolor{black}{, the authors
focus on the data offloading problem in the space-air-ground network
to guarantee the balance between energy consumption and mean time
cost. \cite{436-ultra-dense-LEO} investigate}s the LEO satellite
access networks by integrating with terrestrial network to realize
seamless global communication services. In \cite{1269-jisijing-Mag},
the authors investigate the distributed mobility management framework
of the space-terrestrial networks, via reconfiguration of mobility
management functions, to improve the handover decision efficiency.
In \cite{1271-ultra-dense-LEO-mag}, the authors analyze the communication
system of LEO satellites to investigate the application utilizing
stochastic geometry. \cite{1276-role-LEO} investigates the LEO satellite
constellation to improve the Internet connections for remote areas.
\cite{1278-LEO-mag-zhouhaibo} presents a multi-layer management structure
composed of MEO and LEO satellites to achieve efficient mobility and
resource control.

As for the MEC and NFV in SAGIN, \cite{1273-xuelin-JSAC-LEO} designs
a LEO satellite enabled heterogeneous MEC framework, as well as an
offloading scheme to improve the computational performance. In \cite{JZY-TWC},
a software defined LEO satellite framework as well as VNF deployment
model are proposed, and efficient deployment algorithms are designed.
\cite{1259-caoxuelin-SIN-Mag} proposes a reconfigurable intelligent
surfaces based MEC framework of space information networks to improve
both the communication and computing capabilities. In \cite{1267-megaLEO-CM},
the authors design a LEO satellite aided edge computing platform to
guarantee the computing continuum. \cite{1274-LEO-MEC} proposes a
framework of orbital edge computing with LEO satellite constellations,
to satisfy the growing demand of multiple applications. \cite{1275-sat-computing-FL}
analyzes a serious of possible ways to combine machine learning and
satellite networks to provide satellite based computing. 

\begin{table*}
\centering\caption{\label{tab:SatHAPUAV}Platforms in the space and air.}

\begin{tabular}{|c|>{\centering}m{3.3cm}|>{\centering}m{3.3cm}|>{\centering}m{3.3cm}|}
\hline 
\multirow{2}{*}{Classification} & \textbf{Space} & \multicolumn{2}{c|}{\textbf{Air}}\tabularnewline
\cline{2-4} 
 & Satellite & HAP/Airship  & UAV\tabularnewline
\hline 
Height & $\geq$160km  & 17km-22 km & 600m-18km \tabularnewline
\hline 
Time duration  & Years & Months to years  & Minutes to hours \tabularnewline
\hline 
Coverage & Large and periodic  & Medium and fixed  & Small and dynamic \tabularnewline
\hline 
Energy provider & Solar panel & Solar panel or battery  & Battery \tabularnewline
\hline 
Controller & Ground & Satellite or ground  & Ground\tabularnewline
\hline 
Cost & High & Medium & Low\tabularnewline
\hline 
Applications & Observation

Communication

Navigation 

Astronomy 

Meteorology

\dots{} & Real-time monitoring Communication relay \\
Emergency recovery \\
Rocket launch platform

\dots{}  & Sensing 

Communication \\
Surveillance

Farming

Logistics

\dots{} \tabularnewline
\hline 
Ref & \cite{JZY-TWC,helijun-TMC1,1269-jisijing-Mag} & \cite{ziyejia-jsac,1288-xiaozhenyu-HAPmag} & \cite{1020-dongchao-mag-uav,jzy-IoT-aerialcomputing}\tabularnewline
\hline 
\end{tabular}
\end{table*}

However, from the perspective of LEO satellites, the multi-tier MEC
for SAGIN, as well as the network slicing related researches have
not been thoroughly investigated. Hence, in this work, we explore
the potential of LEO satellite based access patterns in SAGIN, especially
the heterogenous MEC and NFV-based network slicing.

\subsection{Contributions }

In this work, we provide the preliminaries and current development
of SAGIN, and analyze the access patterns from the perspective of
LEO satellites. Besides, advanced techniques such as MEC and NFV are
introduced in the LEO satellites based access networks to promote
the network ability and enhance the service quality, in terms of served
users' number, service diversity, resource flexibility, etc. Both
users and resource providers benefit from the new-type access technologies.
In order to differentiate the general single-layer mobile computing
service, we further present the heterogenous MEC framework as well
as the SFC deployment technique. To clearly elaborate the performance,
use cases are also provided. Moreover, we provide preliminary analyses
on the open issues and possible directions for LEO satellite based
access in SAGIN. Note that in this work, we mainly consider the remote
users and non-terrestrial resources, instead of the abundant urban
wireless communication facilities. The contributions of this work
are summarized as follows:
\begin{itemize}
\item The preliminaries of SAGIN are detailed, in which the LEO satellites
based access patterns are presented, including the employment of advanced
techniques such as MEC, NFV, caching, etc.
\item Based on LEO satellites, the hierarchical MEC as well as SFC implementation
based network slicing for SAGIN are investigated. Besides, corresponding
use cases are provided to validate the proposed schemes.
\item The LEO satellites based access possibilities in SAGIN as well as
open challenges are analyzed, also with promising directions for the
future researches.
\end{itemize}
\par The rest of this work is organized as follows. We firstly provide
an overview of SAGIN in Section \ref{sec:Basics-of-SAGIN}, including
the compositions, resources, demands, applications, etc. In Section
\ref{sec:LEO-advance}, the role and potential of LEO satellites based
access in SAGIN are elaborated, as well as the advances of in-network
computing paradigms MEC and NFV based resource management paradigms,
including validated use cases. The open issues and possible directions
and are presented in Section \ref{sec:Open-Issues}. Finally, we draw
the conclusion in Section \ref{sec:Conclusions}.

\section{Basics of SAGIN \label{sec:Basics-of-SAGIN}}

In this section, we provide the basics of SAGIN including LEO satellites.
Specifically, as illustrated in Fig. \ref{fig:SAGIN}, SAGIN is generally
composed by the hierarchical satellites in the space, multiple vehicles
in the air, and basic infrastructures on the ground. As for the multiple
satellites in the space segment, the non-LEO satellites include GEO
and MEO satellites. For example, the developed GEO/MEO satellites
include MicroGEO \cite{1269-jisijing-Mag}, Boeing, O3B, etc. The
air vehicles include airships, HAPs, aircrafts, and UAVs, on the basis
of different heights and functions. Besides, most users come from
ground, such as emergency areas, ocean, deserts, and remote areas
without services of ground base stations. Note that with the development
of deep space exploration, e.g. lunar exploration and Mars exploration,
the deep space probes are also potential users in SAGIN, since the
explored data should be transmitted back to earth via multi-layer
satellites. In addition, agents such as satellites, HAPs, and UAVs,
act as resources in some cases, and as users in a couple of scenarios.
Note that the meta resources include sensors, transceivers, computations,
caching, etc., equipped on various platforms, to support multiple
applications, such as deep space exploration, air transportation,
ocean service, emergency rescue, remote area users, and navigation
augmentation \cite{1285_LEO_navigation,1277-LEO-mag-positioning}.
Hence, the resource management and efficient allocation are significant
issues, to leverage limited resources to satisfy heterogeneous demands.

The properties of satellites, HAPs, and UAVs are provided in Table
\ref{tab:SatHAPUAV}. Specifically, from the view of height, satellites
in the space operate above 160km, HAPs or airships in the air implement
among 17km-22km, and UAVs in the air have a flight height among 600m-18km,
depending on the detailed demands and platform property. Besides,
satellites can work during many years, HAPs can last for months to
years, and UAVs can continue working as minutes to hours. The coverage
of satellites is large and periodic, HAPs can cover an medium and
fixed area, while UAVs can flexibly cover a small area. From the perspective
of energy provider, satellites utilize the solar panel, HAPs leverage
the solar panels as well as battery to store energy at night, and
lithium batteries are applied in UAVs. In addition, the controllers
for satellites are from ground, HAPs can be controlled by the ground
center or satellites, and the existing UAVs are operated by the ground
controller. The typical applications of satellites include observation,
communication, navigation, astronomy, meteorology, etc. HAPs can support
applications such as real-time monitoring, communication relay, emergency
recovery, and rocket launch platforms. UAVs are applied in the areas
of sensing, communication, surveillance, farming and logistics.

\begin{table*}
\centering\caption{\label{tab:Sat}Satellite preliminaries}
\begin{tabular}{|c|c|c|c|c|c|c|}
\hline 
Satellites & Orbit altitude & Orbit period & Launch cost & Round-trip delay & Network complexity & User switching\tabularnewline
\hline 
\hline 
LEO & 500km-2000km & 1.5h-2h & Low & 30-50ms & High & Frequent\tabularnewline
\hline 
MEO & 2000km-20,000km & 2h-24h & Medium & 125ms-250ms & Medium & Medium\tabularnewline
\hline 
GEO & 20,000km-36,800km & 24h & High & 400ms-600ms & Low & Not necessary\tabularnewline
\hline 
\end{tabular}
\end{table*}

\begin{figure*}
\centering

\includegraphics[scale=0.48]{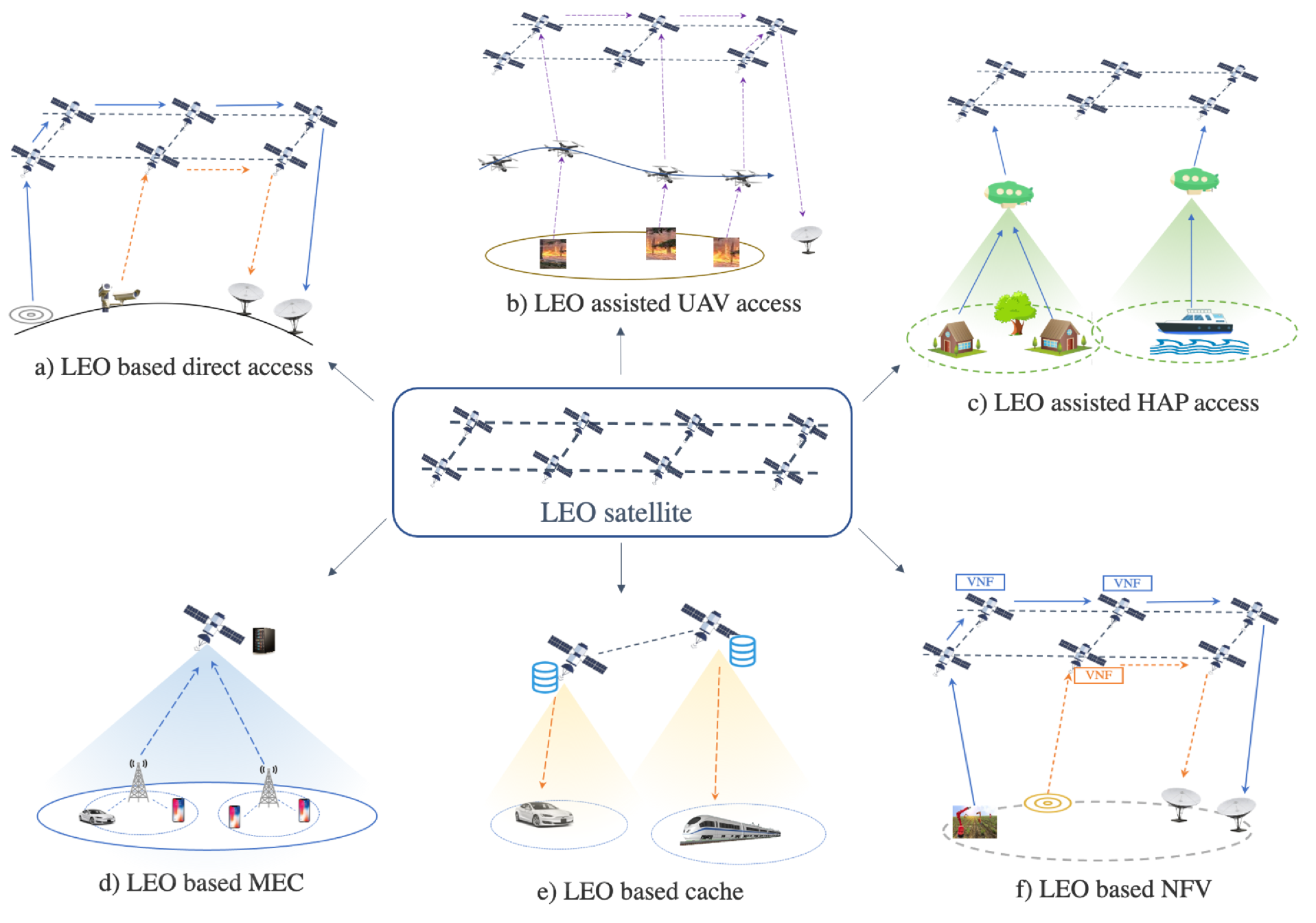}\textcolor{black}{\caption{\label{fig:Access-examples}Access examples from the perspective of
LEO satellites.}
}
\end{figure*}

\section{LEO Satellite based Access\label{sec:LEO-advance}}

In this section, the characteristics of LEO satellite based access
\textcolor{black}{patterns} in SAGIN are analyzed firstly, and further
we introduce the network advances for enhancement, i.e. the hierarchical
MEC as well as the network slicing techniques. Besides, correlated
use cases are provided to verify the corresponding technique implementation.

\subsection{\textcolor{black}{Properties of LEO Satellites}}

LEO satellites attract more attentions from the world due to the networking
cooperation via inter-satellite links and mega-constellation, such
as Starlink of SpaceX and Kuiper of Amazon, to provide abundant applications
for increasing demands from remote IoT, maritime voyage, etc. To make
it clear, in Table \ref{tab:Sat}, by revisiting the properties of
multiple satellites, the potential of LEO satellites is discussed.
In particular, the height of LEO satellites ranges from 500km to 2000km,
the height of MEO satellites is among 2000km to 20,000km, and GEO
satellites operate between 20,000km and 36,800km. Besides, the height
larger than 36,800km is deemed as outer space. The orbit periods of
LEO, MEO, and GEO satellites are 1.5h-2h, 2h-24h, and 24h, respectively.
As for the launch cost, LEO satellites are the lowest. In addition,
the round-trip delay of LEO satellites is 30ms-50ms, while it is 125ms-250ms
and 400ms-600ms for the MEO and GEO satellites, respectively. Hence,
compared with MEO and GEO satellites, LEO satellites can provide lower-delay
services for the terrestrial users. Besides, LEO satellite constellations
dramatically improve the global coverage, including the polar areas.
However, the network complexity of LEO satellites is high, since many
LEO satellites are networking, and serious in the mega-constellation,
while the network complexity of GEO satellite is low since it can
operate dependently to cover 1/3 globe. Indeed, users suffer from
the frequent handover within the coverage of LEO satellites, while
it is not even necessary to switch due to the large coverage and quasi-static
property of GEO satellites. 

From the perspective of LEO satellites, users may come from deep space,
space, air, and ground (including ocean). Moreover, LEO satellites
act as multi-hop relays or servers. In short, the properties and advantages
of LEO satellites reveal why most countries in the world pay close
attentions to LEO network constructions. Accordingly, the challenges
such as network complexity and frequent switch as discussed above
should be addressed.

\subsection{Advances for Enhancement}

The emerged advances such as MEC, and NFV based network slicing techniques,
enable LEO satellites access to flexibly accommodate more users, and
efficiently leverage the heterogenous resources. Fig. \ref{fig:Access-examples}
provides a couple of access examples, from the perspective of LEO
satellites, such as the LEO based direct access, LEO assisted UAV
access, LEO assisted HAP access, etc. Besides, the advanced techniques
such as MEC, cache, and NFV are also applied in the LEO satellites
enabled access network to enhance performance. Note that other than
the direct offloading from terrestrial users to LEO satellites in
the example references \cite{1268-helijun-TMC2,1273-xuelin-JSAC-LEO},
the LEO satellites enabled multi-layer MEC mechanism is more applicable.
Besides, the LEO based slicing in SAGIN is prospective, instead of
independent LEO satellites based VNF deployment in Fig. \ref{fig:Access-examples}.
\textcolor{black}{More specifically, MEC can be deemed as the node-level
resource virtualization technique, while SFC implementation is the
network-level virtualization technique, i.e., network slicing. Accordingly},
the LEO satellite enabled hierarchical MEC as well as the network
slicing are elaborated in the following.

\subsubsection{LEO Satellites Enabled Hierarchical MEC in SAGIN}

\begin{figure}
\centering

\includegraphics[scale=0.36]{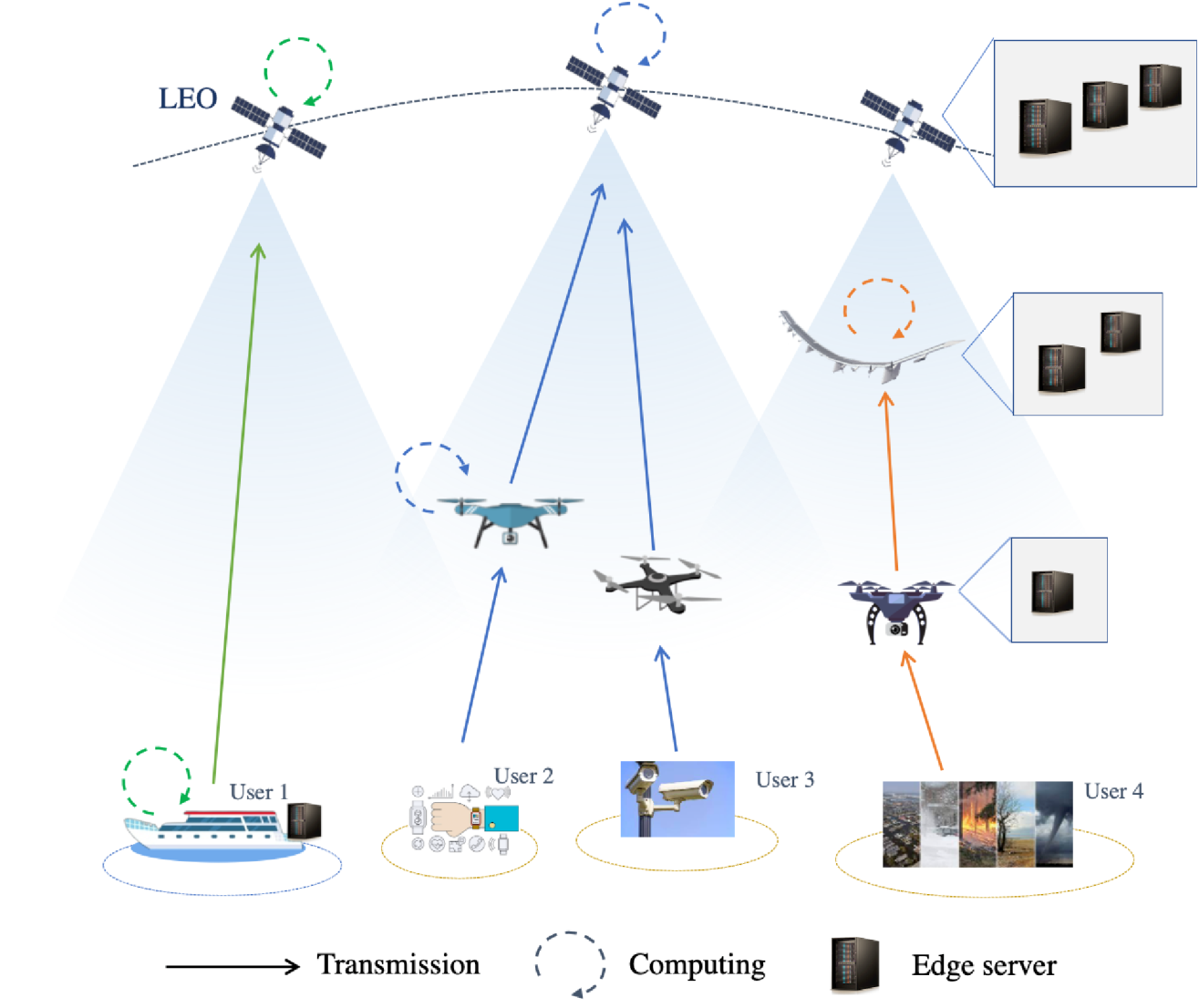}

\textcolor{black}{\caption{\label{fig:MEC}LEO satellite based MEC in SAGIN. }
}
\end{figure}

The MEC paradigm can provide efficient and effective mechanism to
assist terrestrial users to deal with the computation demands. In
SAGIN, satellites, HAPs, UAVs, as well as ground users, are equipped
with different computation capability according to limited load capacity.
The multi-layer MEC enables the computation continuum. As illustrated
in Fig. \ref{fig:MEC}, the terrestrial terminal (user 1) is equipped
with limited computation server capability, so the task is partly
computed locally by user 1, and partly offloaded to the LEO satellite
for online computing. As for terrestrial user 2 and user 3, the task
data are offloaded to the nearby UAVs, and the computation task of
user 2 is partially completed by the UAV. However, due to the limited
loading capacity of UAVs, a portion of data of user 2 is offloaded
to the LEO satellite for further computation. The data from user 3
is offloaded the satellite relayed by a UAV, while the data from user
4 is transmitted by a UAV to the HAP for computation. Note that the
data from user 2 and user 3 share the computation resources of the
same LEO satellite.

\begin{figure}
\centering

\includegraphics[scale=0.44]{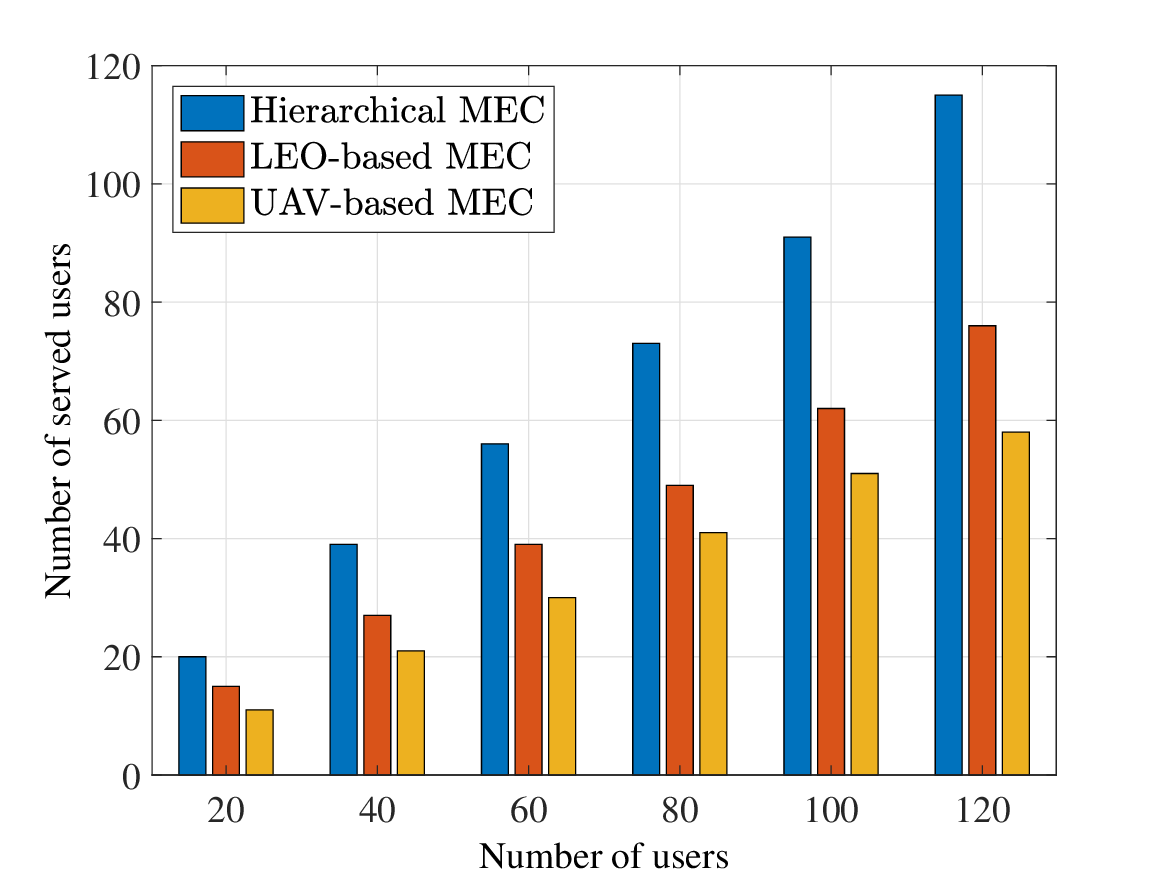}\caption{\textcolor{black}{\label{fig:MEC-simu}LEO based hierarchical MEC
performance.}}
\end{figure}

In particular, according to \cite{1267-megaLEO-CM,jzy-IoT-aerialcomputing},
there exists the following computation offloading modes: binary and
partial. The binary mode means the data is fully offloaded or completed
locally, as the data offloading mode of user 3 and user 4 in Fig.
\ref{fig:MEC}. The partial mode means the cooperation of different
platforms to complete the same computation task, such as the service
procedures of user 1 and user 2 in Fig. \ref{fig:MEC}. Indeed, the
detailed offloading decision policy should be determined according
to optimization objective and considered metrics such as limited energy
supply, communication resource limitation, delay QoS, etc. The general
methods such as game theory, convex optimization, and machine learning
are available. 

\subsubsection*{Numerical example}

We explore the performance of different MEC paradigms in Fig. \ref{fig:MEC-simu}.
Specifi\textcolor{black}{cally, the scenario is set with four LEO
satellites with orbit height of 1000km, four UAVs are uniformly distributed
in a fixed area with altitude of 2km, and 20-120 users are within
the coverage. The hierarchical MEC is implemented by the cooperation
of UAVs and LEO satellites, while the compared methods are single-layer
LEO satellites and single-layer UAVs based MEC.} From Fig. \ref{fig:MEC-simu},
it is observed that with respect to the number of served users, the
hierarchical MEC is outstanding compared to the single-layer MEC,
and the result is in accordance with intuitive understanding. Besides,
the LEO-based MEC performs better that the UAV-based MEC due to the
larger coverage and stronger computing load capacity of LEO satellites.

\subsubsection{LEO Satellites Based Network Slicing in SAGIN}

\textcolor{black}{Recently, due to the advantages such as programmability
and virtualization introduced by NFV and SDN, the related network
slicing techniques are introduced into satellite networks, aiming
at cost-effective resource utilization and high QoS performance, as
well as lowering the capital and operation expenditures \cite{JZY-TWC}.
The basics depend on the advanced reconfigurable resources of both
software and hardware. SDN is equipped with the feature of data and
control separation, and operating in a centralized management mode.
The specific implementation is utilizing NFV technique} to provide
multiple services for different users via decoupling the VNFs from
physical platforms. In addition, a \textcolor{black}{series} of VNFs
for one task form a SFC.\textcolor{black}{{} Moreover, the network slicing
can also be expanded in the SAGIN, instead of only satellite networks.
Such characteristics facilitate an elastic SAGIN, and the SFC deployment
in the SAGIN is a significant mode for efficient resource implementation. }

\begin{figure}
\centering

\includegraphics[scale=0.44]{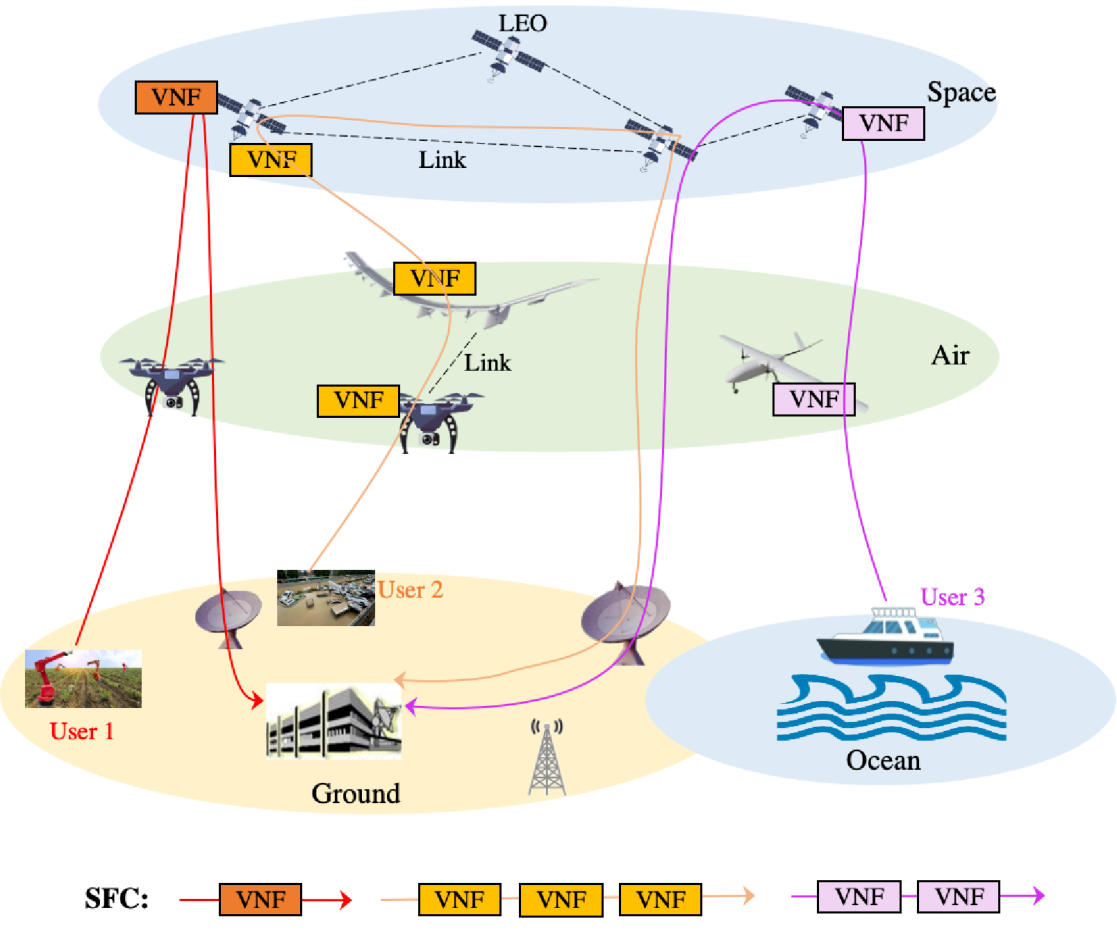}

\caption{\label{fig:SFC} Network slicing via SFC deployment in SAGIN.}
\end{figure}

For clarity, in Fig. \ref{fig:SFC}, a case of netwo\textcolor{black}{rk
slicing scenario via SFC deployment in SAGIN is presented. In particular,
LEO satellites in the space and the aerial vehicles can provide flexible
services leveraging SDN and NFV techniques, and the network slicing
can be realized via SFC deployment in the SAGIN. In detail, there
exists three different SFC requirements in Fig. \ref{fig:SFC}: the
SFC of user 1 includes only one VNF and it is deployed on a LEO sa}tellite,
and the UAV serves as a relay. The SFC corresponding to user 2 is
composed of three VNFs, which are sequently deployed on a UAV, a HAP,
and a LEO satellite, respectively. The SFC from the ocean user 3 includes
two VNFs, which are respectively mapped on a UAV and a LEO satellite.
Note that the SFCs of user 1 and user 2 share the resources of the
same LEO satellite.

The network slicing in SAGIN improves the resource efficiency, for
example, increasing stable coverage, reducing latency, etc. Fundamentally,
the related problem should consider SFC deployment optimization, multiple
resource limitation, the coupled restriction between VNF mapping and
routing selection, as well as time horizon related scheduling should
be handled.

\subsubsection*{Numerical example}

\begin{figure}
\centering

\includegraphics[scale=0.44]{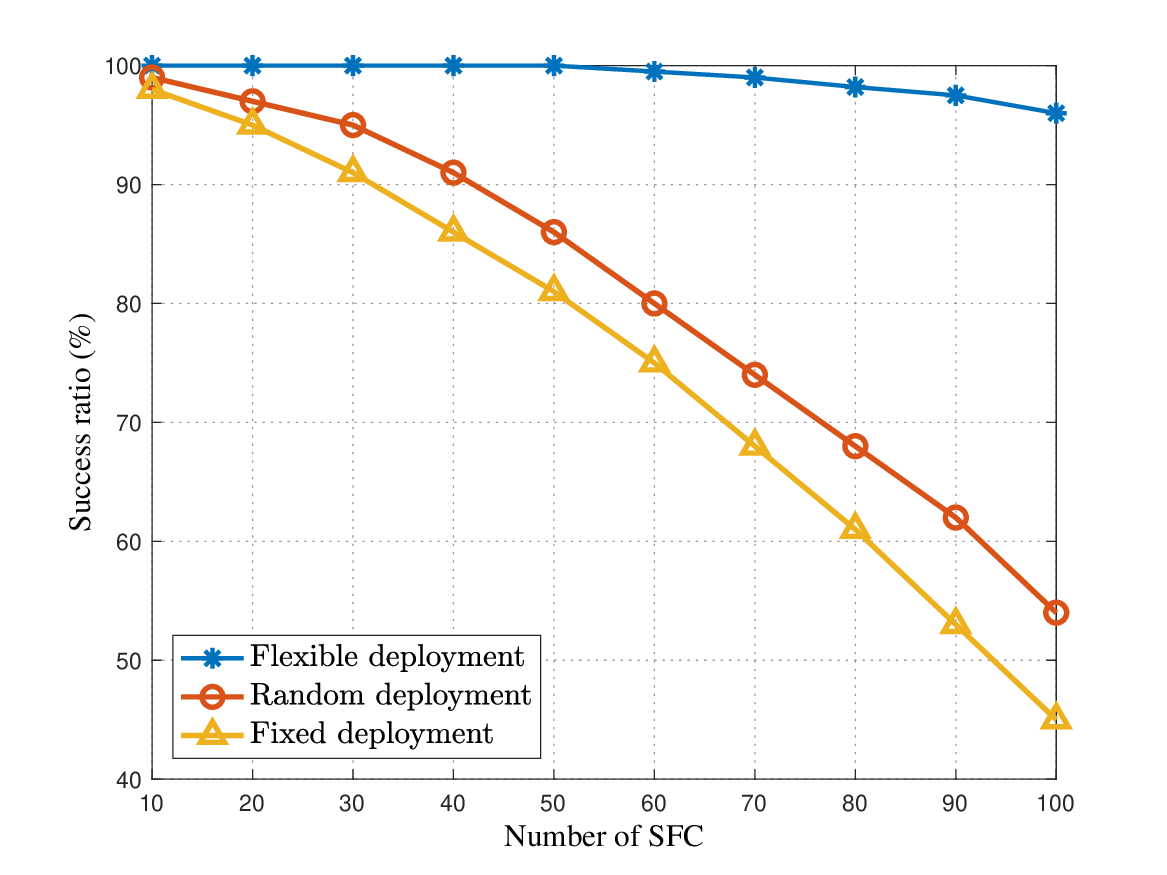}\caption{\textcolor{black}{\label{fig:SFC-simu}Performance of LEO based SFC
implementation. }}
\end{figure}

To clearly verify the performance of LEO satellites enabled network
slicing in SAGIN, the numerical results are conducted in a simple
use case and Fig. \ref{fig:SFC-simu} shows the results. Specifically,
\textcolor{black}{the scenario is set within a small scale Walker
constellation composed of sixteen LEO satellites }with orbit height
of 1000km, six UAVs are uniformly distributed in a fixed area with
altitude of 2km, and 10-100 tasks with SFC requirements. In Fig. \ref{fig:SFC-simu},
different paradigms including flexible deployment, random deployment,
and fixed deployment are compa\textcolor{black}{red. It is obviously
that the flexible SFC deployment performs best, while the random deployment
leads to multiple conflicts and resource inefficiency, which typically
results in SFC failure. As for the fixed deployment, it means NFV
is not supported in such a strategy, and VNFs should be deployed on
specified platforms, so the SFC request is liable to fail if there
is no active communication links.}

\section{Open Issues and Directions\label{sec:Open-Issues}}

\begin{table*}
\centering

\caption{\label{tab:Open-Challenges}Open challenges and possible directions
of LEO-access in SAGIN.}

\begin{tabular}{|>{\centering}p{5cm}|>{\centering}p{5cm}|>{\centering}p{5cm}|}
\hline 
Open challenges & Opportunities & Directions\tabularnewline
\hline 
\hline 
Large number of LEO satellites of mega-constellation & Co-design of various LEO satellite constellations & Interoperability among different constellations via protocol design\tabularnewline
\hline 
High dynamics and frequent users' handover & Cooperation of multiple platforms for continuum & Mobility management, distributed manage system\tabularnewline
\hline 
On-board processing and control & Intelligent platform & Online learning, distributed control\tabularnewline
\hline 
Information safety & Privacy protection & Differential privacy, blockchain\tabularnewline
\hline 
\end{tabular}
\end{table*}

Although LEO satellites open up significant potentials in the SAGIN
access networks, there still exist a couple of open issues, such as
large number of LEO satellites in the mega-constellation, high dynamics
leading to frequent handover, on-board processing and control, information
safety, etc. For clarity, we summarize such issues in Table \ref{tab:Open-Challenges}.
To tackle these challenges, we provide the following possible directions
and tips for future researches.
\begin{itemize}
\item As for the large number of LEO satellites in the mega-constellation,
the opportunities rely on the co-design of various LEO satellite constellations,
and the future directions may advocate increasing the interoperability
among different LEO satellites from different companies or countries
via protocol design. 
\item In order to deal with the high dynamics of LEO satellites, and the
related frequent handover issues of users, the cooperation of multiple
platforms can guarantee continuum to some extent. Besides, the future
directions may depend on the mobility management strategy design,
efficient handover decisions, and developing the distributed manage
system. 
\item To figure out the challenge of on-board processing and control, typically
the intelligent and machine learning based techniques can be employed.
In particular, the online learning mechanism design as well as the
distributed control paradigms may be promising approaches for future
directions.
\item Due to the global service and openness of SAGIN, especially LEO satellite
networks, the information safety correspondingly becomes intractable
issues. Associated with the opportunity of privacy protection, the
emerged popular techniques such as differential privacy as well as
blockchain can be considered as future directions for SAGIN safety.
\end{itemize}
\par Essentially, the open challenges of the LEO satellites access
in SAGIN are not only in terms of the discussed issues as above, the
orbit scarcity, frequency competition, and resource ability limitation
are all issues should be focused, and available strategies should
be devised.

\section{Conclusions \label{sec:Conclusions}}

\textcolor{black}{In this work, we have reviewed the basics and details
of LEO satellites based access modes and techniques in SAGIN. Firstly
we have elaborated the SAGIN preliminaries and current development,
as well as the possible applications. Then, the potentials of LEO
access patterns, including the direct access, multi-hop, relay, caching,
MEC, and NFV implementation have been elaborated. Besides, based on
the recent advanced techniques MEC and NFV, the multi-layer MEC as
well as the SFC deployment network slicing in SAGIN have been detailed,
and corresponding use cases have been verified. Further, the open
challenges and possible directions have also predicted and analyzed.
}We expect this study can promote the development of LEO multiple
access possibilities in SAGIN.

\end{document}